\documentclass[aps,prl,twocolumn,floats,epsfig,showpacs]{revtex4-1}

\usepackage{bbm}
\usepackage{amssymb}
\usepackage{ifpdf}
\usepackage{color}
\usepackage{graphicx,epsfig}

\newcommand{\Z}{{\mathbb{Z}}}

\newcommand{\C}{{\mathbb{C}}}

\begin{document}

\title{Mass and Charge of the Quantum Vortex in the $(2+1)$-d $O(2)$ Scalar 
Field Theory}
\author{M.\ Hornung, Joao C.\ Pinto Barros, and U.-J.\ Wiese}
\affiliation{Albert Einstein Center, Institute for Theoretical Physics, 
University of Bern, Sidlerstrasse 5, 3012 Bern, Switzerland}

\begin{abstract}

Using numerical simulations, a vortex is studied in the broken phase of the
$(2+1)$-d $O(2)$-symmetric scalar field theory in the vicinity of the 
Wilson-Fisher fixed point. The vortex is an infraparticle that is surrounded by 
a cloud of Goldstone bosons. The $L$-dependence of the vortex mass in a 
finite $C$-periodic volume $L^2$ leads to the determination of the renormalized 
vortex charge.

\end{abstract}

\maketitle

Vortices are topological excitations that arise in superfluids, 
superconductors, and Bose-Einstein condensates 
\cite{Abr57}. In three spatial dimensions vortices are line-defects, including 
cosmic strings \cite{Nie73}, that sweep out a world-sheet during their 
time-evolution, while in two dimensions vortices are point-defects. In the 2-d 
classical XY model, they drive the
Berezinskii-Kosterlitz-Thouless phase transition \cite{Ber70,Kos73}. Popov was 
first to note that vortices and phonons in a $(2+1)$-d superfluid are dual to 
charged particles and photons in scalar QED \cite{Pop72}. He concluded that the 
mass of a vortex corresponds to its rest energy divided by the square of the 
speed of sound. Duan found the mass to diverge logarithmically with the volume, 
but attributed a finite mass to the vortex core \cite{Dua94}. According to Baym 
and Chandler, the core-mass corresponds to the mass of the superfluid within 
the core \cite{Bay83}. An equivalent concept, the Kopnin-mass exists for 
superconductors and fermionic superfluids \cite{Kop78,Kop91,Kop98,Vol98}. 
Thouless and Anglin studied the reaction of a vortex to an external force by a 
pinning potential by means of the Gross-Pitaevskii equation \cite{Tho07}. They 
confirmed that the vortex mass receives an infinite contribution. A recent 
Monte Carlo study in the $(2+1)$-d $O(2)$ model, using boundary conditions that 
break translation invariance, concluded that the vortex mass is finite in the 
infinite volume limit \cite{Del19}, thus contraditing the previous results. A 
finite vortex mass in $(2+1)$-d is also inconsistent with the fact that vortex 
loops and global cosmic strings in the $(3+1)$-d $O(2)$ model, which are again 
surrounded by massless Goldstone bosons, have a tension that increases 
logarithmically with the string length \cite{Vil85,Bra87}.

\begin{figure}[tbp]
\includegraphics[width=0.5\textwidth]{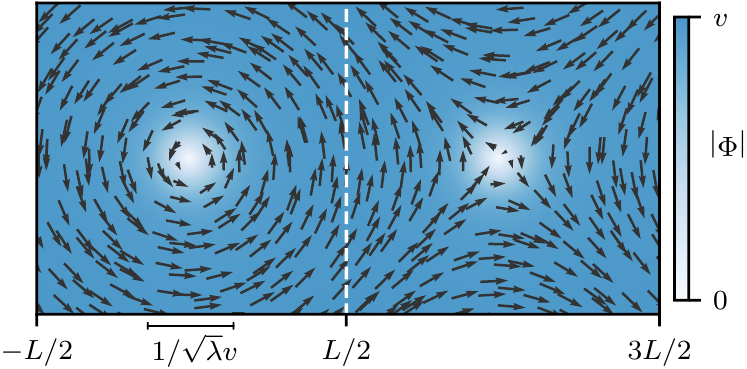}
\caption{\textit{Vortex field in a C-periodic volume $L^2$, together 
with its twisted charge conjugation copy in the neighboring square.}}
\label{Fig1}
\end{figure}

Here we consider the problem fully non-perturbatively from first principles. 
A deep insight from algebraic quantum field theory, which we apply here to
condensed matter physics, is that, due to the soft cloud of phonons surrounding 
it, the vortex is a non-local infrared sensitive infraparticle 
\cite{Sch63,Fro82,Buc82,Buc82a,Buc14} (like a charged particle). By enclosing 
it in a translation invariant C-periodic volume (characterized by a charge 
conjugation twist) \cite{Pol91}, we carry out a fully controlled Monte Carlo 
calculation which shows that the vortex mass is indeed logarithmically 
divergent. Due to its infraparticle nature, a moving vortex and a vortex at 
rest do not belong to the same superselection sector \cite{Buc82}. Hence,
its kinetic mass need not agree with its rest mass, which is confirmed by our 
numerical study. A C-periodic vortex field is illustrated in Fig.\ref{Fig1}.

Vortices arise as classical solutions in the broken phase of the $(2+1)$-d 
$O(2)$ symmetric field theory for a complex scalar field $\Phi(x) \in \C$ with 
the Lagrangian
\begin{equation}
{\cal L} = \frac{1}{2} \partial_\mu \Phi^* \partial^\mu \Phi - V(\Phi), \
V(\Phi) = \frac{\lambda}{4!}\left(|\Phi|^2 - v^2\right)^2.
\end{equation}
Static classical vortex solutions have the form 
$\Phi(r,\varphi) = f(r) \exp(i \varphi)$ and obey the Euler-Lagrange equation
\begin{eqnarray}
&&\Delta \Phi = \frac{\lambda}{6}\left(|\Phi|^2 - v^2\right)\Phi \ \Rightarrow
\nonumber \\
&&\left(\partial_r^2 + \frac{1}{r} \partial_r - \frac{1}{r^2}\right) f =
\frac{\lambda}{6}\left(f^2 - v^2\right) f.
\end{eqnarray} 
At large distances $\Phi$ approaches the vacuum value $v$ of the scalar field 
as $f(r) \sim v - 3/(\lambda v r^2)$. The energy density 
${\cal H} = \frac{1}{2} \vec \nabla \Phi^* \cdot \vec \nabla \Phi + V(\Phi)$ of 
the static vortex is
\begin{equation}
{\cal H}(r) = \frac{1}{2} \left(\left(\partial_r f\right)^2 + 
\frac{f^2}{r^2}\right) + \frac{\lambda}{4!}\left(f^2 - v^2\right)^2 \sim 
\frac{v^2}{2 r^2}.
\end{equation}
This leads to an infrared logarithmic divergence of the vortex mass. 
Integrating the energy density over a disc of radius $R$ one obtains
\begin{equation}
E(R) = 2 \pi \int_0^R dr \ r {\cal H}(r) \ \Rightarrow \
E(R) \sim \pi v^2 \log\frac{R}{R_0}.
\end{equation}
The logarithmic divergence is due to a cloud of massless Goldstone bosons that
surrounds the vortex. As we will see, the quantum vortex is dual to a charged
particle in $(2+1)$-d scalar QED, with the Goldstone boson being the dual
photon. The divergence of the vortex mass arises because the logarithmic 
Coulomb potential in two spatial dimensions is confining. In this way, the
prefactor $e^2/(4 \pi) = \pi v^2$ of the logarithm is associated with the dual 
electric charge $e$ of the vortex.

The semi-classical treatment of vortices is limited to the quantization of 
their collective degrees of freedom. Here, for the first time, we present a 
completely controlled, fully non-perturbative, translation invariant 
calculation of the quantum vortex in the continuum limit of the 
$(2+1)$-d $O(2)$ scalar field theory, approaching the Wilson-Fisher fixed point 
from the broken phase. For this purpose, we regularize the theory on a 3-d 
cubic Euclidean space-time lattice. The complex scalar field is represented by 
a unit-vector $(\cos(\varphi_x),\sin(\varphi_x))$ associated with the lattice 
sites $x$. The resulting $(2+1)$-d XY model is defined by the partition function
\begin{equation}
Z = \prod_x \frac{1}{2 \pi} \int_{-\pi}^\pi \!\!\!\!
d\varphi_x \exp(- S[\varphi]), \ 
S[\varphi] = \sum_{\langle xy\rangle} s(\varphi_{xy}).
\end{equation}
Here $x$ and $y$ are nearest-neighbor lattice points and 
$\varphi_{xy} = \varphi_x - \varphi_y$. The standard action has 
$s(\varphi_{xy}) = \frac{1}{g^2} (1 - \cos\varphi_{xy})$, while the Villain 
action \cite{Vil75} is given by 
\begin{equation}
\exp(- s(\varphi_{xy})) = \!\! \sum_{n_{xy} \in \Z} \!\! 
\exp\left(- \frac{1}{2 g^2} (\varphi_{xy} - 2 \pi n_{xy})^2\right).
\end{equation}

The XY model can be dualized exactly to a $(2+1)$-d Abelian gauge theory with 
integer-valued non-compact vector potentials $A_l \in  2 \pi \Z$ associated with
the links $l$ of the dual lattice. The dual partition function is
\begin{equation}
Z = \prod_l \sum_{A_l \in 2 \pi \Z} \exp(- S[A]), \ 
S[A] = \sum_\Box \widetilde{s}(F_\Box).
\end{equation}
Here the field strength $F_\Box = d A_l = A_{l_1} + A_{l_2} - A_{l_3} - A_{l_4} = 
2 \pi n_\Box$ is the lattice curl of the vector potentials associated with the 
four links $l_1$, $l_2$, $l_3$, and $l_4$ that encircle a dual plaquette $\Box$.
The dual Boltzmann weight is given by
\begin{equation}
\exp(- \widetilde{s}(F_\Box))\!=\!\frac{1}{2 \pi} \int_{-\pi}^\pi \!\!\!\!
d\varphi_{xy} \exp(- s(\varphi_{xy})\!+\! i n_\Box \varphi_{xy}), 
\end{equation}
where $\Box$ is the plaquette dual to the original nearest-neighbor link 
$\langle xy \rangle$. The dual standard action is given by the modified Bessel
function $\exp(- \widetilde{s}(F_\Box)) = I_{n_\Box}(1/g^2)$, while the dual 
Villain action is simply given by 
$\widetilde{s}(F_\Box) = \frac{1}{2 e^2} F_\Box^2$. Here $e$ is the dual charge 
that obeys the Dirac quantization condition $e g = 2 \pi$.

Next, we relate the dual integer gauge theory to $(2+1)$-d QED with a charged 
scalar field $\chi_{\widetilde x} \in U(1)$ that represents the vortex in the dual
description. Here $\widetilde x$ is a dual lattice site (the center of a cube 
on the original lattice). The corresponding lattice action is given by
\begin{eqnarray}
\label{QEDaction}
&&S[A,\chi] = \frac{1}{2 e^2} \sum_\Box F_\Box^2 - \varkappa \sum_l 
\mbox{Re}\left(\chi^*_{\widetilde x} \exp(i A_l) \chi_{\widetilde y}\right) \
\Rightarrow \nonumber \\
&&S[A,\chi = 1] = \frac{1}{2 e^2} \sum_\Box F_\Box^2 - \varkappa \sum_l \cos A_l.
\end{eqnarray}
Here $l$ is the link that connects the dual nearest-neighbor sites 
$\widetilde x$ and $\widetilde y$. The action is gauge invariant against
\begin{equation}
A_l' = A_l - d\alpha_{\widetilde x} =
A_l + \alpha_{\widetilde x} - \alpha_{\widetilde y}, \
\chi_{\widetilde x}' = \exp(i \alpha_{\widetilde x}) \chi_{\widetilde x}. 
\end{equation}
Here $d$ is the lattice gradient. In the unitary gauge, $\chi_{\widetilde x} = 1$,
the action reduces to the second line of eq.(\ref{QEDaction}). Taking the limit 
$\varkappa \rightarrow \infty$ leads to the constraint $A_l \in 2 \pi \Z$, 
which corresponds to the integer gauge theory that is dual to the $(2+1)$-d XY 
model. Scalar QED exists in two phases. The Coulomb phase at large values of 
$e$ is dual to the broken phase of the $(2+1)$-d XY model at small values of 
$g$, with the Goldstone boson being the dual photon. The charged particle in 
the Coulomb phase of $(2+1)$-d scalar QED is dual to the vortex in the 
$(2+1)$-d XY model, with $e$ being the bare vortex charge. At small values of 
$e$, the theory exists in a Higgs phase in which the photon picks up a mass. 
This phase, in which vortices condense, is dual to the massive symmetric phase 
of the $(2+1)$-d XY model at large values of $g$. 

Here we concentrate on the Coulomb phase in which vortices are dual to charged 
scalar particles with bare charge $e$. A charged particle is surrounded by a 
cloud of massless photons, which extends to infinity. The resulting non-local 
object is known as an infraparticle, which, due to the soft photon cloud, does 
not simultaneously have a well-defined charge and a well-defined mass 
\cite{Buc82}. The operator that creates the infraparticle is given by
\begin{equation}
\label{vortexfield}
\chi_{\widetilde x}^C = \exp\left(i \alpha_{\widetilde x}^C\right) \chi_{\widetilde x} =
\exp\left(i \Delta^{-1} \delta A_l\right) \chi_{\widetilde x}.
\end{equation}
It leads from the vacuum into the charge 1 super\-selection sector. The operator
is non-local because it contains not only $\chi_{\widetilde x}$, but also the
non-local Coulomb cloud surrounding the charge, which is represented by
$\alpha_{\widetilde x}^C = \Delta^{-1} \delta A_l$. Here $\Delta$ is the 2-d 
spatial Laplacian, and $\delta A_l = A_{l_1} + A_{l_2} - A_{l_3} - A_{l_4}$ is the 
2-d lattice divergence of $A_l$, which is constructed from the links $l_1$ and
$l_2$ that exit the dual site $\widetilde x$ in the positive 1- and 2-direction,
and the links $l_3$ and $l_4$ that enter $\widetilde x$ from the negative 1- 
and 2-direction. In fact, $\alpha_{\widetilde x}^C$ is the gauge transformation
that turns the gauge field $A_l$ into the Coulomb gauge 
\begin{equation}
\delta A_l' = \delta (A_l - d\alpha_{\widetilde x}^C) = 
\delta A_l - \Delta \alpha_{\widetilde x}^C = 0 \ \Rightarrow \
\alpha_{\widetilde x}^C = \Delta^{-1} \delta A_l.
\end{equation}
Here we have used $\delta d \alpha_{\widetilde x}^C = 
\Delta \alpha_{\widetilde x}^C$. The gauge transformation 
$\exp(i \alpha_{\widetilde x}^C)$ endows the charged particle with its surrounding
Coulomb field. Under gauge transformations
\begin{equation}
{\alpha_{\widetilde x}^C}' = \Delta^{-1} \delta A_l' = 
\Delta^{-1} \delta A_l - \Delta^{-1} \delta d\alpha_{\widetilde x} = 
\alpha_{\widetilde x}^C - \alpha_{\widetilde x},
\end{equation}
such that the scalar field $\chi_{\widetilde x}^C$ in the Coulomb gauge is
gauge invariant and represents the non-local charged particle, which is just 
dual to the vortex surrounded by a cloud of massless Goldstone bosons.

The physical charged field from above was first constructed by Dirac
\cite{Dir55} and was also used by Fr\"ohlich and Marchetti in their construction
of monopole super\-selection sectors in the Coulomb phase of $(3+1)$-d compact 
$U(1)$ lattice gauge theory \cite{Fro86}. In that case, the monopoles are dual 
to massive charged infraparticles in the Coulomb phase of $(3+1)$-d scalar QED.
Fr\"ohlich and Marchetti have also provided fully non-perturbative 
constructions of soliton sectors in a wide variety of systems \cite{Fro87}. 
This includes vortices in the Higgs phase of 3-d scalar QED which become anyons 
in the presence of a Chern-Simons term \cite{Fro88,Fro89}. The vortices 
considered here are dual to the charged particles in the Coulomb phase of 
$(2+1)$-d scalar QED. Because in two spatial dimensions the Coulomb potential 
is confining, as we already discussed at the classical level, the mass of an 
isolated vortex diverges logarithmically in the infrared.

While the vortex sectors in the physical Hilbert space are removed to infinite
energy in the infinite volume limit, it is most interesting to construct them
in a finite volume. A recent study that enforced a vortex by boundary 
conditions and fitted a vortex profile to numerical data concluded that the
vortex mass has a finite infinite volume limit \cite{Del19}. This contradicts 
our results. For an infrared sensitive infraparticle, imposing fixed boundary 
conditions (which break translation invariance) can be problematical. Periodic 
boundary conditions maintain translation invariance, but, as a consequence of 
Gauss' law, they do not allow the existence of charged states. For this reason, 
C-periodic boundary conditions were introduced both for Abelian \cite{Pol91} 
and for non-Abelian gauge theories \cite{Kro91,Wie92,Kro93}. When shifted by a 
distance $L$, a C-periodic field is replaced by its charge-conjugate. 
C-periodic boundary conditions leave translation invariance intact and allow 
the existence of charged states. However, charge and anti-charge states are 
mixed to form charge conjugation eigenstates. C-periodic boundary conditions 
are also used in lattice simulations of monopoles \cite{Pol91,Jer99} and of QCD 
coupled to QED \cite{Luc16}. C-periodic boundary conditions in scalar QED are 
discussed in detail in the supplementary material.

We consider $(2+1)$-d scalar QED in the unitary gauge, $\chi_{\widetilde x} = 1$, 
at $\varkappa = \infty$ such that $A_l \in 2 \pi \Z$. We work on a cubic 
space-time lattice with periodic temporal and C-periodic spatial boundary 
conditions, $A_{l'} = {^CA_l} = - A_l$. The link $l'$ is shifted relative to $l$ 
by a distance $L$ in the spatial 1- or 2-direction. C-periodic Abelian gauge 
fields are anti-periodic, because charge conjugation changes the sign of the 
vector potential. In the unitary gauge the vortex field of 
eq.(\ref{vortexfield}) takes the form 
$\chi_{\widetilde x}^C = \exp\left(i \Delta^{-1} \delta A_l\right)$. With 
C-periodic boundary conditions the spatial Laplacian $\Delta$ has no zero-modes 
and $\Delta^{-1} \delta A_l$ is well-defined. 

There is a $\Z(2)$ symmetry that changes the sign of the vortex field 
$\chi_{\widetilde x}^C$, which characterizes a non-trivial superselection sector 
and guarantees the stability of the vortex in a C-periodic volume. This symmetry
results from a constant gauge transformation $\alpha_{\widetilde x} = \pi$, such
that $\exp(i \alpha_{\widetilde x}) = - 1$. This is the only global gauge 
transformation that is consistent with C-periodic boundary conditions. We 
denote it as vortex field reflection.

In a C-periodic volume the real-part $\mbox{Re}\chi_{\widetilde x}^C$ (which is 
C-even) obeys periodic while the imaginary part $\mbox{Im}\chi_{\widetilde x}^C$ 
(which is C-odd) obeys anti-periodic boundary conditions. This implies that the 
C-even and C-odd components of the charged vortex state necessarily have 
different spatial momenta and thus different energies. Since C-periodic 
boundary conditions maintain translation invariance, we construct vortex fields
with definite spatial momenta $(p_1,p_2)$ at fixed Euclidean time 
${\widetilde x}_3$
\begin{eqnarray}
&&\chi^+(p_1,p_2,{\widetilde x}_3) = \sum_{{\widetilde x}_1,{\widetilde x}_2}
\exp(i p_1 {\widetilde x}_1 + i p_2 {\widetilde x}_2) 
\mbox{Re}\chi_{\widetilde x}^C, \nonumber \\
&&\chi^-(p_1,p_2,{\widetilde x}_3) = \sum_{{\widetilde x}_1,{\widetilde x}_2}
\exp(i p_1 {\widetilde x}_1 + i p_2 {\widetilde x}_2) 
\mbox{Im}\chi_{\widetilde x}^C.
\end{eqnarray}
The momenta of the periodic C-even component $\chi^+(p_1,p_2,{\widetilde x}_3)$ 
are quantized in integer units, $p_i = 2 \pi n_i/L$, $n_i \in \Z$, while the 
momenta of the anti-periodic C-odd component $\chi^-(p_1,p_2,{\widetilde x}_3)$
are quantized in half-odd-integer units, $p_i = 2 \pi (n_i + \frac{1}{2})/L$. 
Specifically, we consider $\chi^+(0,0,{\widetilde x}_3)$ at zero momentum and 
$\chi^-(p_1,p_2,{\widetilde x}_3)$ at the smallest possible momenta 
$p_i = \pm \pi/L$. At large Euclidean time separation, the corresponding 
correlation functions then decay exponentially
\begin{eqnarray}
\label{vortexcorrelators}
&&\left\langle \chi^+(0,0,0) \chi^+(0,0,{\widetilde x}_3) 
\right\rangle \sim \exp(- m {\widetilde x}_3), \nonumber \\
&&\left\langle \chi^-(p_1,p_2,0) \chi^-(p_1,p_2,{\widetilde x}_3)^*
\right\rangle \sim \exp(- E {\widetilde x}_3).
\end{eqnarray}
Here $m$ is the rest mass of the C-even component of a vortex, while $E$ is the 
energy of the C-odd component that moves with minimal momentum 
$p_i = \pm \pi/L$. Both components are odd under vortex field reflection. 

First, we have simulated the $(2+1)$-d lattice XY model in the broken phase
using the Wolff cluster algorithm \cite{Wol89}, applied both to the standard 
and Villain action. The model then has massless Goldstone bosons that are 
described by a low-energy effective field theory with the Euclidean action
$S[\varphi] = 
\int d^3x \frac{\rho}{2} \partial_\mu \varphi \partial_\mu \varphi$.
We have performed numerical simulations in various space-time volumes in the
vicinity of the critical point, in order to determine the spin stiffness $\rho$ 
which we use to set the energy scale. Near the critical point, physical 
quantities $O$ with mass dimension 1 (including $\rho$, $m$, and $E$) scale as
\begin{equation}
\label{scaling}
O(1/g^2) \! = \! A_O (1/g_c^2 \! - \! 1/g^2)^\nu 
\left[1 \! + \! a_O (1/g_c^2 \! - \! 1/g^2)^\omega\right].
\end{equation} 
Here $\nu = 0.67169(7)$ is the very accurately known critical exponent
associated with the correlation length, and $\omega = 0.789(4)$ is 
a universal exponent that controls corrections to scaling \cite{Has19}. The 
amplitude $A_O$ and the coefficient $a_O$ are observable-specific and not 
universal. However, the amplitude ratios $A_m/A_\rho$ and $A_E/A_\rho$ are 
universal and yield the continuum limit results approaching the $O(2)$ 
symmetric Wilson-Fisher fixed point from the broken phase. Our results for 
$\rho$ are consistent with those of \cite{Neu03}. The critical coupling for the 
standard action, $1/g_c^2 = 0.4541652(11)$, \cite{Cam06} and for the Villain 
action, $1/g_c^2 = 3.00239(6)$ \cite{Neu03} are known very accurately.


\begin{figure}[tbp]
\includegraphics[width=0.5\textwidth]{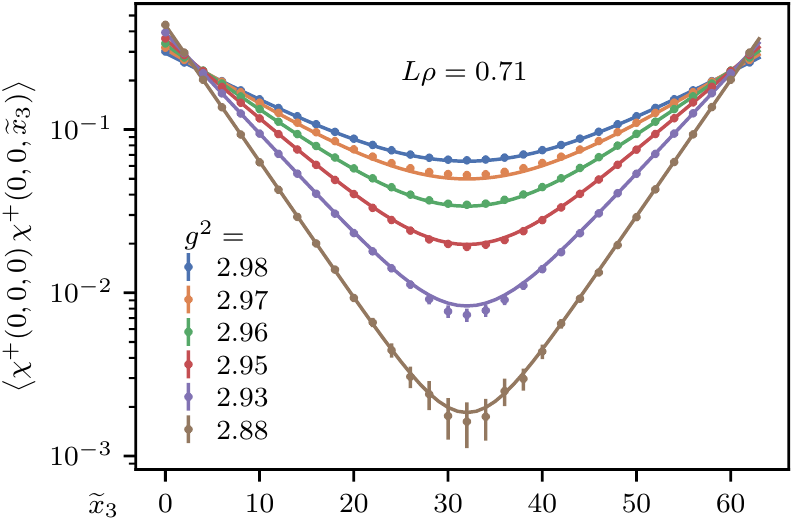}
\caption{\textit{Cosh-fits of
$\left\langle \chi^+(0,0,0) \chi^+(0,0,{\widetilde x}_3) \right\rangle$.}}
\label{Fig2}
\end{figure}

\begin{figure}[tbp]
\includegraphics[width=0.5\textwidth]{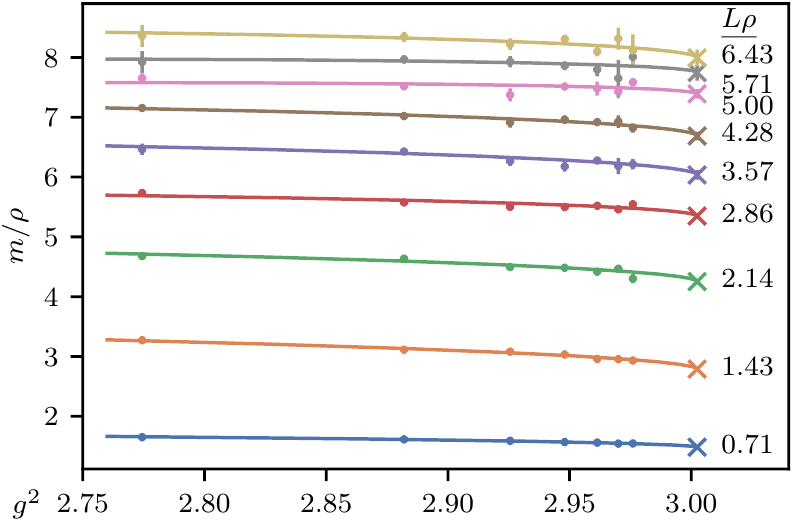}
\caption{\textit{Continuum extrapolation of the vortex mass in units of the 
spin stiffness, $m/\rho$, for different spatial sizes $L \rho$.}}
\label{Fig3}
\end{figure}

Next, we have simulated the vortex correlation functions of 
eq.(\ref{vortexcorrelators}) in the dual $(2+1)$-d $\Z$ gauge theory using a 
Metropolis algorithm. The quantities $m$ and $E$ are extracted from cosh-fits 
of the correlation functions at large Euclidean time separations 
${\widetilde x}_3$, as illustrated for the Villain action in Fig.\ref{Fig2}. In 
order to take the continuum limit, we approach the critical point and increase 
the number of lattice points while keeping the physical size $L \rho$ fixed. We 
then take the continuum limit of the dimensionless ratios $m/\rho$ and $E/\rho$ 
by identifying the universal amplitude ratios $A_m/A_\rho$ and $A_E/A_\rho$ from 
fits to the scaling form of eq.(\ref{scaling}), as illustrated for the Villain 
action in Fig.\ref{Fig3}. The continuum limit extrapolations for the standard 
action are consistent within statistical errors, thus confirming universality
when approaching the fixed point.

\begin{figure}[tbp]
\includegraphics[width=0.5\textwidth]{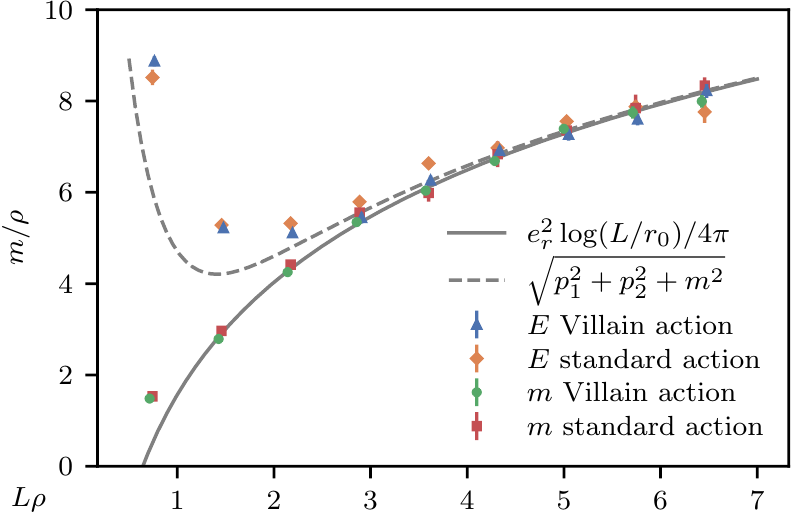}
\caption{\textit{Vortex mass $m$ and energy $E$ as functions of $L \rho$ in 
the continuum limit. The dashed curve shows $\sqrt{p^2 + m^2}$ for the momenta 
$p_1,p_2 = \pm \frac{\pi}{L}$, which differs from $E$.}}
\label{Fig4}
\end{figure}

The final continuum limit results for $m/\rho = A_m/A_\rho$ and 
$E/\rho = A_E/A_\rho$ are shown in Fig.\ref{Fig4} as a function of the spatial 
size in physical units $\rho L$. For large $L$, the mass of the quantum vortex 
diverges logarithmically, 
\begin{equation}
m \sim \frac{e_r^2}{4 \pi} \log(L/r_0), \ e_r = 2 \sqrt{3.58(8) \pi \rho},
\label{logscaling}
\end{equation}
thus confirming the expected Coulombic confinement that we already encountered 
for the classical vortex. The prefactor of the logarithm determines the 
renormalized vortex charge $e_r$, which is another universal feature of the 
Wilson-Fisher fixed point. As expected, in the large volume limit, the energy 
$E$ of the (then more and more slowly moving) C-odd component approaches the 
mass $m$ of the C-even component (which has zero momentum). 

The energy $E$ differs from the relativistic expression $\sqrt{p^2 + m^2}$, 
since an infraparticle breaks Lorentz invariance spontaneously \cite{Buc82}, in 
addition to the explicit breaking due to the finite volume. We define the 
kinetic finite-volume mass $m_k$ of the vortex as
\begin{equation}
E = m + \frac{p^2}{2 m_k}, \ p_1, p_2 = \pm \frac{\pi}{L} \Rightarrow 
m_k = \frac{\pi^2}{L^2 (E - m)}. 
\end{equation}
For $L \rho = 1.43(2)$ and $2.14(3)$, we obtain $m_k/m = 0.71(3)$ and 
$0.55(4)$, indicating significant differences between the kinetic and the rest 
mass. The typical size of the vortex core can be characterized by 
$r_0 = 0.64(3)/\rho$ in eq.(\ref{logscaling}), or by its universal charge 
radius (yet to be determined).

Taking the infraparticle nature of the vortex into account sheds light on the 
subtle concepts of its rest and kinetic mass. It would be most interesting to 
investigate the universal properties of the quantum vortex in experiments with 
superfluid films or Bose-Einstein condensates tuned to the vicinity of the 
Wilson-Fisher fixed point, e.g., by extracting the universal vortex charge 
$e_r$ from the Coulomb interactions of vortices or anti-vortices.

\newpage

The research leading to these results has received funding from the 
Schweizerischer Na\-tio\-nal\-fonds.


\begin{thebibliography}{99}

\bibitem{Abr57}
A.\ A.\ Abrikosov, Zh.\ Eksp.\ Teor.\ Fiz.\ 32 (1957) 1442;
Sov.\ Phys.\ JETP 5 (1957) 1174.

\bibitem{Nie73}
H.\ B.\ Nielsen, P.\ Olesen, Nucl.\ Phys.\ B61 (1973) 45.

\bibitem{Ber70}
V.\ L.\ Berezinskii, Sov.\ Phys.\ JETP 32 (1970) 493.

\bibitem{Kos73}
J.\ M.\ Kosterlitz, D.\ J.\ Thouless, J.\ Phys.\ C6 (1973) 1181.

\bibitem{Pop72} 
V.\ N.\ Popov, Sov.\ Phys.\ JETP 37 (1972) 341.

\bibitem{Dua94} 
J.-M.\ Duan, Phys.\ Rev.\ B49 (1994) 12381.

\bibitem{Bay83}
G.\ Baym, E.\ Chandler, J.\ Low Temp.\ Phys.\ 50 (1983) 57.

\bibitem{Kop78}
N.\ B.\ Kopnin, JETP Lett.\ 27 (1978) 390.

\bibitem{Kop91}
N.\ B.\ Kopnin, M.\ M.\ Salomaa, Phys.\ Rev.\ B44 (1991) 9667.

\bibitem{Kop98}
N.\ B.\ Kopnin, V.\ M.\ Vinokur, Phys.\ Rev.\ Lett.\ 81 (1998) 3952.

\bibitem{Vol98} 
G.\ E.\ Volovik, JETP Lett.\ 67 (1998) 528.

\bibitem{Tho07}
D.\ J.\ Thouless, J.\ R.\ Anglin, Phys.\ Rev.\ Lett.\ 99 (2007) 105301.


\bibitem{Del19}
G.\ Delfino, W.\ Selke, A.\ Squarcini, Phys.\ Rev.\ Lett.\ 122 (2019) 050602.

\bibitem{Vil85}
A.\ Vilenkin, Phys.\ Rep.\ 121 (1985) 263.

\bibitem{Bra87}
R.\ H.\ Brandenberger, Int.\ J.\ Mod.\ Phys.\ A2 (1987) 77.






\bibitem{Sch63}
B.\ Schroer, Fortschr.\ Phys.\ 173 (1963) 1527.

\bibitem{Fro82}
J.\ Fr\"ohlich, G.\ Morchio, F. Strocchi, Phys.\ Lett.\ B89 (1979) 61.

\bibitem{Buc82}
D.\ Buchholz, K.\ Fredenhagen, Commun.\ Math.\ Phys.\ 84 (1982) 1.

\bibitem{Buc82a}
D.\ Buchholz, Commun.\ Math.\ Phys.\ 85 (1982) 49.

\bibitem{Buc14}
D.\ Buchholz, J.\ E.\ Roberts, Commun.\ Math.\ Phys.\ 330 (2014) 935.

\bibitem{Pol91}
L.\ Polley, U.-J.\ Wiese, Nucl.\ Phys.\ B356 (1991) 629.

\bibitem{Vil75}
J.\ Villain, J.\ Phys.\ 36 (1975) 581.

\bibitem{Dir55}
P.\ A.\ M.\ Dirac, Canad.\ J.\ Phys.\ 33 (1955) 650.

\bibitem{Fro86}
J.\ Fr\"ohlich, P.\ A.\ Marchetti, Euro.\ Phys.\ Lett. 2 (1986) 933.

\bibitem{Fro87}
J.\ Fr\"ohlich, P.\ A.\ Marchetti, Commun.\ Math.\ Phys.\ 112 (1987) 343.

\bibitem{Fro88}
J.\ Fr\"ohlich, P.\ A.\ Marchetti, Lett.\ Math.\ Phys.\ 16 (1988) 347.

\bibitem{Fro89}
J.\ Fr\"ohlich, P.\ A.\ Marchetti, Commun.\ Math.\ Phys.\ 121 (1989) 177.

\bibitem{Kro91}
A.\ S.\ Kronfeld, U.-J.\ Wiese, Nucl.\ Phys.\ B357 (1991) 521.

\bibitem{Wie92}
U.-J.\ Wiese, Nucl.\ Phys.\ B375 (1992) 45.

\bibitem{Kro93}
A.\ S.\ Kronfeld, U.-J.\ Wiese, Nucl.\ Phys.\ B401 (1993) 190.

\bibitem{Jer99}
J.\ Jersak, T.\ Neuhaus, H.\ Pfeiffer, Phys.\ Rev.\ D60 (1999) 054502.

\bibitem{Luc16}
B.\ Lucini, A.\ Patella, A.\ Ramos, N.\ Tantalo, JHEP 1602 (2016) 076.

\bibitem{Wol89}
U.\ Wolff, Phys.\ Rev.\ Lett.\ 62 (1989) 361.

\bibitem{Has19}
M.\ Hasenbusch, Phys.\ Rev.\ B100 (2019), 224517.

\bibitem{Cam06}
M.\ Campostrini, M.\ Hasenbusch, A.\ Pelissetto, E.\ Vicari, 
Phys.\ Rev.\ B74 (2006) 144506.

\bibitem{Neu03}
T.\ Neuhaus, A.\ Rajantie, K.\ Rummukainen, Phys.\ Rev.\ B67 (2003) 014525.



\end{thebibliography}
\end{document}


\title{Supplementary Information: \\ 
Mass and Charge of the Quantum Vortex in the $(2+1)$-d $O(2)$ Scalar 
Field Theory}
\author{M.\ Hornung, Joao C.\ Pinto Barros, and U.-J.\ Wiese}
\affiliation{Albert Einstein Center, Institute for Theoretical Physics, 
University of Bern, Sidlerstrasse 5, 3012 Bern, Switzerland}

\maketitle

\section{C-periodic boundary conditions}

In scalar QED in the continuum C-periodic boundary conditions imposed on a 
square-shaped spatial volume $L \times L$ take the form
\begin{eqnarray}
&&A_\mu(x + L e_i) = {^CA_\mu}(x) - \partial_\mu \varphi_i(x) = 
- A_\mu(x)  - \partial_\mu \varphi_i(x), \nonumber \\
&&\chi(x + L e_i) = \exp(i \varphi_i(x)) {^C\chi}(x) = 
\exp(i \varphi_i(x)) \chi(x)^*, 
\end{eqnarray}
where $e_i$ is the unit-vector that points in the spatial $i$-direction,
and $\varphi_i(x)$ is a transition function that represents a gauge 
transformation at the boundary. In the Euclidean time 3-direction, periodic
boundary conditions are imposed over an extent $\beta$ that sets the inverse
temperature
\begin{eqnarray}
&&A_\mu(x + \beta e_3) = A_\mu(x) - \partial_\mu \varphi_3(x), \nonumber \\
&&\chi(x + \beta e_3) = \exp(i \varphi_3(x)) \chi(x).
\end{eqnarray}
The transition functions $\varphi_\mu(x)$ are physical degrees of freedom of the
gauge field which ensure that the gauge variant fields $A_\mu(x)$ and $\chi(x)$
are periodic in Euclidean time and $C$-periodic in space only up to gauge
transformations. Combining shifts in various space-time dimension, one derives
the following cocycle consistency conditions
on the transition functions
\begin{eqnarray}
&&\varphi_j(x + L e_i) - \varphi_i(x) = \varphi_i(x + L e_i) - \varphi_j(x) +
2 \pi n_{ij}, \nonumber \\
&&\varphi_3(x + L e_i) + \varphi_i(x) = \varphi_i(x + \beta e_3) - \varphi_3(x) +
2 \pi n_{i3}. \nonumber \\ \,
\end{eqnarray}
Here $n_{\mu\nu} = - n_{\nu\mu} \in \Z$ with $n_{i3} = n_{j3}$ is known as the 
twist tensor. Performing a general (not necessarily periodic or C-periodic)
gauge transformation, $A_\mu(x)' = A_\mu(x) - \partial_\mu \alpha(x)$, one
obtains the gauge transformation behavior of the transition functions
($n_\mu \in \Z$).
\begin{eqnarray}
&&\varphi_i(x)' = \varphi_i(x) + \alpha(x + L e_i) + \alpha(x) - 2 \pi n_i,
\nonumber \\
&&\varphi_3(x)' = \varphi_3(x) + \alpha(x + \beta e_3) - \alpha(x) - 2 \pi n_3.
\end{eqnarray}
The twist tensor then transforms as 
\begin{equation}
n_{ij}' = n_{ij} + 2 n_i - 2 n_j, \ n_{i3}' = n_{i3} - 2 n_3.
\end{equation}
By gauge transformations, the twist tensor can be restricted to 
$n_{\mu\nu} \in \{0,1\}$. 

\section{Duality in a finite volume with twisted 
$C$-periodic boundary conditions}

The duality transformation that relates the vortex to a dual charged particle 
is typically discussed in the infinte volume.
It is, however, possible to perform it in a finite volume with twisted
$C$-periodic boundary conditions in space and periodic boundary
conditions in Euclidean time. Analogous to the fields and their duals, the 
boundary conditions of the original and the dual theory are then related by a 
Fourier transform.

We start the dualization by recalling the partition function of the dual 
integer gauge theory,
\begin{equation}
	Z = \left( \prod_{l} \sum_{A_{l} \in 2 \pi \mathbb{Z}} 
	\right)\prod_{\square}\exp(-\tilde{s}(F_{\square})) \, .
\end{equation}
In the first step of the dualization, the dual
weights $ \exp(-\widetilde{s}(F_{\square}))$ are
expressed by their Fourier representations as
\begin{align}
Z &=\prod_{l} \sum_{A_{l} \in 2 \pi \mathbb{Z}} 
\prod_{\square} \int_{-\pi}^{\pi} d\eta_{\square}
\nonumber	\\
&\quad \times  \prod_{\square} \exp(-s(\eta_{\square})) \exp(-i\eta_{\square} 
F_{\square})\, .
\end{align}
The newly introduced auxiliary field $\eta$ is attached to the plaquettes of
the dual lattice, or equivalently to the links $\left<xy\right>$ of the 
original lattice. Next a partial integration is performed. Here the boundary
conditions need to be taken into account. A somewhat lengthy but
straightforward calculation yields 
\begin{equation}
\label{eq:twistedpartial}
\sum_{\square}\eta_{\square}F_{\square} = \sum_{l}\delta \eta_{\square} A_{l}  
+ P_1 n_{23}  + P_2 n_{31} - P_3n_{12} \, .
\end{equation}
Here $\delta \eta $ is the lattice curl of $\eta$ evaluated on the original
lattice. It is attached to the links $l$ of the  dual lattice or,
equivalently, to the plaquettes of the original lattice. Furthermore, 
$n_{\mu \nu}$ is the twist tensor characterizing the boundary conditions of the 
dual gauge theory. The $P_\mu$ correspond to the Polyakov loops of $\eta$,
\begin{equation}
P_\mu = \sum_{\square \in \sigma_\mu} \eta_{\square} \, .
\end{equation}
Here $\sigma_\mu$ is a set of plaquettes dual to links that form a closed loop 
on the original lattice,  which wraps around the boundary in the 
$\mu$-direction. The constraint $\delta \eta_{\square}=0$, which will soon 
arise, ensures that $P_\mu$ is independent of the form of the path. There is an 
additional term in eq.\ (\ref{eq:twistedpartial})
involving the transition functions that characterize the boundary
conditions for the gauge theory. It vanishes due to the constraint $\delta
\eta_{\square}=0$, and is therefore omitted.  The field $\eta$ is subject to
$C$-periodic boundary conditions in space and periodic boundary conditions in
Euclidean time, without a twist. Now the sum over $A$ is performed, resulting
in the aforementioned constraint on $\eta$,
\begin{multline}
Z= \left( \prod_{\square}\int_{-\pi}^{\pi}d\eta_{\square} \right)
\prod_{l}\delta(\delta \eta_{\square}) \prod_{\square} \exp(-s(\eta_{\square})) \\
\exp(-i( P_1 n_{23} + P_2 n_{31} - P_3 n_{12} ) )\, .
\end{multline}
In the final step of the dualization, we switch from the dual lattice to the
original lattice. The auxiliary field $\eta$ is then attached to the links 
$\left<xy\right>$, and the constraint $\delta \eta_\square=0$ corresponds to 
$d \eta_{xy}=0$. Furthermore, the constrained integration over $\eta_{xy}$ is 
replaced by an unconstrained integration over the spin field $\varphi_x$, which 
is attached to the lattice sites and obeys twisted $C$-periodic boundary 
conditions in space and twisted periodic boundary conditions in time.
In the infinite volume, this procedure is
justified by the Poincar\'e lemma which ensures that all closed forms are exact.
Here we show in a finite volume with $C$-periodic boundary conditions in space
and periodic boundary conditions in Euclidean time, that for each $\eta_{xy}$ 
with $d\eta_{xy}=0$ there exists a field $\varphi_x$ and boundary twists
$\theta_\mu$ for it, such that $\eta_{xy} =  \varphi_{xy} = \varphi_{y} -
\varphi_{x}$.

\begin{figure}[htpb]
\centering
\includegraphics{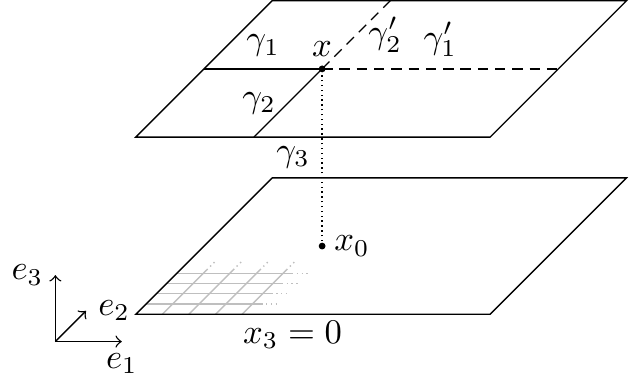}
\caption{Illustration of the paths $\gamma_\mu$ and $\gamma_\mu^{\prime}$ 
employed in the construction of the field $\varphi_x$ with 
$\eta_{xy}= \varphi_{xy}= \varphi_{y}-\varphi_x$.}%
\label{fig:gammas}
\end{figure}

Consider the three definitions 
\begin{align}
\varphi_x &= \frac{1}{2}(\theta_1 + \sum_{\left<yz\right> \in \gamma_1}\eta_{yz}
- \sum_{\left<yz\right> \in \gamma_1'} \eta_{yz}) \, , \nonumber \\
\varphi_x &= \frac{1}{2}(\theta_2 + \sum_{\left<yz\right> \in \gamma_2}\eta_{yz}
- \sum_{\left<yz\right> \in \gamma_2'} \eta_{yz}) \, \nonumber \\
\varphi_x &= \varphi_{x_0} +
\sum_{\left<yz\right>\in\gamma_3}\eta_{yz} \, ,
\end{align}
with spatial boundary twists $\theta_1,\theta_2$ that obey 
$\theta_1 - \theta_2= P_1-P_2$ and a temporal boundary twist of $\theta_3=P_3$. 
The sets of links $\gamma_i$ and $\gamma_i'$ correspond to the paths 
illustrated in Figure~\ref{fig:gammas}.
By construction it follows that
$\eta_{xy}= \varphi_{xy}= \varphi_y-\varphi_x$. Furthermore, the constraint 
$d\eta_{xy} = 0$ ensures that the three definitions of $\varphi_x$ coincide for 
all $x$. The same constraint also implies that $P_1-P_2=0 \mod \pi$ and 
$P_3 = 0 \mod \pi$.

This construction is not unique, and there is a remaining global $O(2)$
symmetry. Consider $\varphi_x$ and $\overline{\varphi}_x$, such that 
$\varphi_{xy} = \overline{\varphi}_{xy} = \eta_{xy}$.
On the links within the volume, that do not cross a boundary, this implies that
they differ by a constant, $\varphi_x-\overline{\varphi}_x=c$. For links that 
cross a twisted $C$-periodic boundary, this implies that 
$\theta_i-\overline{\theta}_i =2 c$
and for a link that crosses the (possibly twisted) periodic temporal boundary,
it implies that $\theta_3-\overline{\theta}_3 =0$. The construction of
$\varphi_x$ is therefore unique only up to the following global $O(2)$ symmetry,
\begin{equation}
\varphi_x' = \varphi_x+\delta,\qquad \theta_i'= \theta_i +\delta,\qquad 
\theta_3'=\theta_3 .
\end{equation}
We can therefore replace the constrained integration over the auxiliary link
variables $\eta_{xy}$ by an unconstrained integration over the spin variables 
$\varphi_x$. The final form of the partition function reads 
\begin{multline}
Z = \frac{1}{2\pi}\left(\prod_{x}\int_{-\pi}^{\pi}d\varphi_{x}  
\int_{-\pi}^{\pi} d\theta_1 \sum_{\theta_2 \in \{ \theta_1,\theta_1+\pi\}} 
\sum_{\theta_3 \in \{0, \pi\}} \right) 
\\ \times \prod_{\left<xy\right>} \exp(-s(\varphi_{xy}))  
\exp(-i (\theta_1-\theta_2 )n_{31} -i \theta_3n_{12} )  \, .
\end{multline}
The division by $2\pi$ is due to the additional redundancy because of the
global $O(2)$ symmetry.  It is now apparent that the gauge invariant 
characteristic of the boundary conditions of the dual gauge theory, the twist 
tensor, is related to the boundary conditions of the original spin model by a 
Fourier transform. 